\documentclass[11pt]{article}
\usepackage{amsmath}
\usepackage{amsfonts}

\def\be{\begin{equation}}
\def\ee{\end{equation}}
\def\bea{\begin{eqnarray}}
\def\eea{\end{eqnarray}}
\def\bean{\begin{eqnarray*}}
\def\eean{\end{eqnarray*}}

\def\hg{\hat{g}}

\def\hOmega{\hat{\Omega}}
\def\hT{\hat{T}}

\def\scri{\mathcal{I}}

\def\hg{\hat{g}}

\def\hG{\hat{G}}
\def\hT{\hat{T}}

\def\hOmega{\hat{\Omega}}
\def\hT{\hat{T}}

\def\tt{\tilde{t}}
\def\ta{\tilde{a}}

\def\tc{\tilde{c}}
\def\tu{\tilde{u}}
\def\tR{\widetilde{R}}
\def\tG{\widetilde{G}}
\def\tnabla{\widetilde{\nabla}}
\def\tx{\tilde{x}}
\def\tE{\tilde{E}}
\def\tC{\widetilde{C}}
\def\tT{\widetilde{T}}
\def\tA{\widetilde{A}}
\def\tL{\widetilde{L}}
\def\tg{\tilde{g}}
\def\tphi{\tilde{\phi}}

\def\tGamma{\widetilde{\Gamma}}

\def\cg{\check{g}}

\def\cT{\check{T}}

\def\cG{\check{G}}

\def\cOmega{\check{\Omega}}

\def\ba{\bf{a}}
\def\bb{\bf{b}}
\def\bc{\bf{c}}
\def\bd{\bf{d}}
\def\bi{\bf{i}}
\def\bj{\bf{j}}
\def\bk{\bf{k}}

\begin{document}

\title{Conformal Methods in Mathematical Cosmology}
\author{Paul Tod\\Mathematical Institute,\\Oxford University,\\ Woodstock Road, Oxford OX2 6GG\\ UK.\footnote{email: tod@maths.ox.ac.uk }}

\maketitle
\begin{abstract}
When he first introduced the notion of a conformal boundary into the study of asymptotically empty space-times, Penrose noted that that the boundary would be null, space-like or time-like according as the cosmological constant $\Lambda$ was zero, positive or negative. While most applications of the idea of a conformal boundary have been to the zero-$\Lambda$, asymptotically-Minkowskian case, there also has been work on the nonzero cases. Here we review work with a positive $\Lambda$, which is the appropriate case for cosmology of the universe in which we live.

\end{abstract}

\section{Introduction}
When he first introduced the notion of a conformal boundary into the study of asymptotically empty space-times, \cite{RP1,RP2}, Penrose noted that the boundary would be null, space-like or time-like according as the cosmological constant $\Lambda$ was zero, positive or negative. While most applications of the idea of a conformal boundary have been to the zero-$\Lambda$, asymptotically-Minkowskian case, there also has been work on the nonzero cases, and I'll review some of that here. We shall concentrate on positive $\Lambda$, which is the appropriate case for cosmology of the universe in which we live.

With a space-like future conformal boundary, we may follow Friedrich \cite{F1,F2} and contemplate using the boundary as a Cauchy surface with data for a conformally-extended set of Einstein equations. We review this material in Section 3. At the other end of the universe, assuming a suitable form of Penrose's Weyl Curvature Hypothesis, \cite{RP4}, we may also contemplate rescaling an initial `Big Bang' singularity for use as a Cauchy surface for another similar set of equations. This will be discussed in Section 4. Finally, in Section 5, we recall Penrose's `outrageous suggestion' \cite{RP3} and contemplate a universe of successive aeons, each of which is an expansion from a rescaled Big Bang surface to a rescaled future conformal boundary which in turn provides the rescaled Big Bang of the next aeon. Now, by assumption, there is a regular conformal metric common to all aeons, while the (conformally-related) physical metrics run from singularity to conformal infinity in each aeon.

\section{Conformal rescaling}

This section is just to fix conventions, which do vary across the literature, though the content should be familiar.

Following \cite{RP2}, I'll be assuming the space-time signature is $(+,-,-,-)$ and the Riemann tensor is defined by
\[(\nabla_a\nabla_b-\nabla_b\nabla_a)V^c=R_{abd}^{\;\;\;\;\;c}V^d.\]

We need to impose Penrose's {\it{asymptotic simplicity}}, or at least {\it{weak asymptotic simplicity}} \cite{RP1,RP2}. Following \cite{RP2}, take the physical metric to be $\tg_{ab}$, the unphysical metric to be $g_{ab}$, and these to be related by conformal rescaling
\be\label{i1}
g_{ab}=\Theta^2\tg_{ab},
\ee
with smooth conformal factor $\Theta$, so that towards the conformal boundary at physical infinity we can take $\Theta\rightarrow 0$. We'll adopt the convention that $\nabla_a,\tnabla_a$ stand for the corresponding metric convariant derivatives. As part of the asymptotic conditions we assume that $\Theta$ has some appropriate degree of smoothness near the surface $\Theta=0$, which is the conformal boundary or $\scri$, and that $d\Theta$ is finite and nonzero there. The Christoffel symbols for the two associated metric connections are related by
\be\label{i2}
\Gamma^a_{bc}-\tGamma^a_{bc}=\delta^a_b\Upsilon_c+\delta^a_c\Upsilon_b-g^{ae}g_{bc}\Upsilon_e,
\ee
where $\Upsilon_a=\Theta^{-1}\Theta_a$ (and of course $g^{ae}g_{bc}=\tg^{ae}\tg_{bc}$).

The Weyl tensors for the two metrics, with indices appropriately arranged,  are equal:
\be\label{i3}
\tC_{abc}^{\;\;\;\;\;d}=C_{abc}^{\;\;\;\;\;d},\ee
but the Ricci tensors have a more complicated relation:

\be\label{i4}\tR_{ab}=R_{ab}-2\nabla_a\Upsilon_b-2\Upsilon_a\Upsilon_b-g_{ab}(\nabla_c\Upsilon^c-2\Upsilon_c\Upsilon^c),\ee
whence also the Ricci scalars are related by
\be\label{i5}\tR=\Theta^2R-6\Theta\Box\Theta+12g^{ab}\Theta_a\Theta_b.\ee
The first key result follows from (\ref{i5}): if the Einstein equations with cosmological constant $\Lambda$ are satisfied by the physical metric then\footnote{With the conventions we are using, the Einstein equations are $\tG_{ab}=-8\pi G\tT_{ab}-\Lambda \tg_{ab}$.}
\[\tR=8\pi G\tT+4\Lambda\]
where $\tT=\tg^{ab}\tT_{ab}$ is the trace of the physical energy-momentum tensor. If $\tT$ goes to zero at the conformal boundary $\scri$, either because it is zero everywhere or because matter terms are decaying ``at infinity" then (\ref{i5}) with finiteness of $R$ and $\Box\Theta$ (implied by smoothness of $g_{ab}$ and $\Theta$) implies
\[g^{ab}\Theta_a\Theta_b=\Lambda/3\mbox{   at }\scri.\]
Thus $\scri$ is time-like, space-like or null according as $\Lambda$ is negative, positive or zero, and furthermore $\Theta_a$ is a suitable normal to $\scri$.

For the next result, from  (\ref{i4}) we obtain that the trace-free part of the tensor
\[2\nabla_a\Theta_b+\Theta(\tR_{ab}-R_{ab})\]
vanishes at $\scri$, so provided the trace-free part of the physical $\tT_{ab}$ is bounded (or, more likely, is tending to zero) at $\scri$ we deduce that the trace-free part of $\nabla_a\Theta_b$ is zero at $\scri$, i.e. $\scri$ is {\emph{umbilic}}. By refining the choice of $\Theta$ one can make $\scri$ extrinsically-flat for nonzero $\Lambda$ or shear-free and expansion-free for zero $\Lambda$.

For the final result in this section we note the transformation
\be\label{i6}\tnabla_d\tC_{abc}^{\;\;\;\;\;d}  =\nabla_dC_{abc}^{\;\;\;\;\;d}-C_{abc}^{\;\;\;\;\;d}\Upsilon_d.\ee
The Bianchi identity gives
\[\tnabla_d\tC_{abc}^{\;\;\;\;\;d}=2\tnabla_{[a}\tL_{b]c},\]
where, assuming the Einstein equations, the Schouten tensor is
\[\tL_{bc}=4\pi G(\tT_{bc}-\frac13\tT\tg_{bc})-\frac16\Lambda \tg_{ab}\]
so that, provided $\tT_{ab}$ decays suitably towards infinity, the left-hand-side in (\ref{i6}) goes to zero at $\scri$. Multiply (\ref{i6}) by $\Theta$ to conclude that
\[C_{abc}^{\;\;\;\;\;d}\Theta_d=0\mbox{  at  }\scri.\]
If $\scri$ is non-null, this at once forces $C_{abc}^{\;\;\;\;\;d}$ to vanish at $\scri$. If $\scri$ is null, more work is required to arrive at the same conclusion. This is a strong result which we shall see again.

\medskip

It's worth noting how the conformal method is particularly welcoming to a trace-free energy momentum tensor. Given a symmetric tensor $\tT_{ab}$ we note that
\[\tg^{ab}\tnabla_a\tT_{bc}=g^{ab}\nabla_a(\Theta^{-2}\tT_{bc})+\tg^{ab}\tT_{ab}\Upsilon_c.\]
Thus if $\tT_{ab}$ is a physical energy-momentum tensor satisfying the conservation equation with respect to $\tnabla$ {\it{and}} is trace-free, then $T_{ab}:=\Theta^{-2}\tT_{ab}$ is also symmetric and trace-free and satisfies the conservation equation with respect to $\nabla$.

\section{Data at $\scri$}

In order to pose an initial value problem with data at $\scri$, one needs a conformally-extended formulation of Einstein's equations for the unphysical metric $g_{ab}$, valid at $\Theta=0$, which implies the usual Einstein equations for the physical metric $\tg_{ab}$ in the region $\Theta>0$. This was achieved by Friedrich \cite{F1} in the vacuum case,  by starting from the identity (\ref{i4}). If we first consider the vacuum equations with cosmological constant then the left-hand-side in (\ref{i4}) is
\[\tR_{ab}=\Lambda\tg_{ab}=\Lambda \Theta^{-2}g_{ab}.\]
Substitute into (\ref{i4}) and multiply by $\Theta^2$ to obtain
\be\label{3.1}\Theta^2R_{ab}-2\Theta^2\nabla_a(\Theta^{-1}\nabla_b\Theta)-2\nabla_a\Theta\nabla_b\Theta-g_{ab}(\Theta^2\nabla_c(\Theta^{-1}\nabla^c\Theta)-2\nabla_c\Theta\nabla^c\Theta+\Lambda)=0.\ee
When differentiated out, all the inverse powers of $\Theta$ cancel and this becomes a non-singular set of second-order PDEs in $(g_{ab},\Theta)$. The challenge now is to get it into a form to which existence theorems can be applied. I'll only sketch this here as the original papers are readily available and straightforward to read.

The idea is to prolong the system to reduce it to a first-order system by introducing more variables and to work in a tetrad formalism. Following \cite{PR} we introduce bold letters for tetrad sub and superscripts. Thus a tetrad can be written $\{e_{\bf a}\}$ for $\ba=0,\ldots,3$ and orthonormality in the unphysical metric is expressed by
\[g(e_{\bf{a}},e_{\bf{b}})=\eta_{\bf{a}\bf{b}}\mbox{,   the Minkowski metric}.\]
The Ricci rotation coefficients for the tetrad are denoted by $\gamma^{\ba}_{\;\bb\bc}$ and are obtained by first-order equations from the tetrad. Likewise the curvature is obtained by first-order equations from the Ricci rotation coefficients. Since (\ref{3.1}) has second-order derivatives of $\Theta$, we introduce a vector $\Sigma_{\ba}:=\nabla_{\ba}\Theta$ to lower the order ($\Sigma_{\ba}$ was called $s_{\ba}$ in \cite{F2}) so these terms become 
$\nabla_{\ba}\Sigma_{\bb}$. Friedrich also introduces the variable $s=\frac14\Box\Theta$.

At the end of section 2 we saw that $C_{abc}^{\;\;\;\;\;d}=0$ at $\scri$ so, assuming smoothness, we can introduce a new (smooth) quantity
\[d_{\ba\bb\bc\bd}:=\Theta^{-1}C_{\ba\bb\bc\bd},\]
and then, by (\ref{i6}), the Bianchi identity for $\tg_{ab}$ can be written
\[\nabla_{\bd}d_{\ba\bb\bc}^{\;\;\;\;\;\bd}=0,\]
which joins the set of first-order equations.

The rest of the unphysical curvature consists of the unphysical Ricci scalar, $R$, and the trace-free part of the unphysical Ricci tensor which Friedrich incorporates in a new quantity:
\[s_{\ba\bb}:=\frac12(R_{\ba\bb}-\frac14R\eta_{\ba\bb}).\] 

Now Friedrich writes out the first-order system for the variables $(e_{\ba}, \gamma^{\ba}_{\;\bb\bc},\Theta,\Sigma_{\ba}, s, s_{\ba\bb},d_{\ba\bb\bc\bd})$. Note that $R$ doesn't appear in this set of variables, though it does appear in the first-order system. This is related to the residual conformal invariance in the equations: if $\Theta$ is replaced by $\widehat\Theta=f\Theta$ for positive $f$ then the variables transform among themselves, and by suitable choice of $f$ any arbitrary choice of $R$ can be made.

\medskip

By  choosing suitable coordinate and tetrad gauge conditions, Friedrich \cite{F1} is able to extract a symmetric hyperbolic system and prove existence and uniqueness in suitable function spaces given suitable data at $\scri$. The data are determined on $\scri$, which can be any 3-manifold, not just ${\mathbb{S}}^3$ as it would be for de Sitter,  by choice of 3-metric, say $a_{\bi\bj}$ where $\bi,\bj,\bk...=1,2,3$, and a second symmetric tensor $c_{\bi\bj}$  (called $d_{\bi\bj}$ in \cite {F1}) satisfying
\[a^{\bi\bj}c_{\bi\bj}=0=D^{\bi}c_{\bi\bj},\]
where $D_{\bi}$ is the Levi-Civita derivative for $a_{\bi\bj}$. The solution exists for a finite interval of the coordinate time used but, since $\scri$ is infinitely far away in the physical metric, this will be an infinite amount of physical time.

In Friedrich's approach,  $\Lambda$ is a constant of integration. With data at $\scri$ (understood as $\scri^+$) close to data for the de Sitter metric, Friedrich's solution may exist all the way back to a past infinity, $\scri^-$, and beyond, possibly through several such `aeons' (where I am pre-empting Penrose's CCC terminology) but these solutions have no nice initial Big Bang. When they do stop, they are likely to have messy initial singularities. It is also implicit in Friedrich's method that one could choose a Cauchy surface inside the space-time and evolve all the way forward to $\scri^+$.

\medskip

In \cite{F2}, Friedrich extended his method to non-vacuum solutions with trace-free energy momentum tensor and dealt in detail with Einstein--Yang-Mills+$\Lambda$  metrics. These have more data at $\scri$: in addition to $(a_{ij},c_{ij})$ there are a Lie-algebra-valued connection one-form which we can write $A^{(\alpha)}_i$, where $(\alpha)$ is the Lie-algebra index, and a vector $E^{(\alpha)}_i$ which is the electric part of the SD Yang-Mills field. Omitting the Yang-Mills index in the interest of clarity, these are subject to the rescaling freedom
\[(a_{ij},c_{ij}, E_i,A_i)\rightarrow(\ta_{ij},\tc_{ij},\tE_i,\tA_i)=(\theta^2a_{ij},\theta^{-1}c_{ij},\theta^{-1}E_i,A_i),\]
and to constraints which, following \cite{F2}, we can give in a space-spinorial form: with spinor equivalents\footnote{now in abstract indices.}
\[\mbox{As spinor fields:   }c_{ij}=E_{ABCD},\;E_i=\phi_{AB},\;A_i=\alpha_{AB},\;D_i=D_{AB}\]
we require
\bean D^{AB}E_{ABCD}&=&2\kappa H<\phi_{A(C}|\phi^{\dagger\; A}_{D)}>\\
2D^C_{\;\;(A}\alpha_{B)C}-[\alpha_{AC},\alpha_B^{\;\;C}]&=&\phi_{AB}+\phi^\dagger_{AB}\\
D^{AB}\phi_{AB}+[\alpha^{AB},\phi_{AB}]&=&0
\eean
to hold on $\scri$, where $<|>$ is the Killing form of the YM gauge group and $[\;\;,\;\;]$ is the Lie-algebra commutator. The second two are what is left of the YM equations.

\medskip

A radiation fluid source can be included in Friedrich's picture \cite{LV}, as can a massless Vlasov source \cite{JTV}. Both these references use conformal methods from a Cauchy surface within the space-time and evolve all the way to $\scri$. More recently, Friedrich has extended his conformal methods to cover some cases of source for which the energy-momentum tensor is not trace-free: some particular massive scalar field sources in \cite{F3}, and dust \cite {F4} (both still with $\Lambda$ of course).

Conformal methods are not essential for this kind of work and there is a large literature investigating long-time existence into the future in the presence of a positive cosmological constant. For example: \cite{HR} proves long-time existence in physical time for Einstein-Vlasov plus  scalar fields with a potential plus $\Lambda$, without conformal methods; \cite{HS,RS,S} do this for various Einstein-Euler plus $\Lambda$ situations; and there are many earlier references, going back at least to \cite{W}, to what was called `the cosmological no-hair conjecture' (or sometimes `theorem'). It is convenient to put all these classes in the context of a {\it{Starobinsky expansion}}.

\medskip

The Starobinsky expansion of the space-time metric in the presence of a positive cosmological constant $\Lambda=3H^2,H>0$ is \cite{st}
\be\label{s1}
g=dt^2-e^{2Ht}h_{ij}dx^idx^j
\ee
where
\be\label{s2}h_{ij}=a_{ij}+e^{-2Ht}b_{ij}+e^{-3Ht}c_{ij}+...
\ee
where all powers $e^{-nHt}$ may appear in $h_{ij}$ except for $n=1$, and the coefficient metrics $a_{ij},b_{ij},c_{ij}...$ are $t$-independent. The coordinate system is defined geometrically: choose a space-like hypersurface to be $t=0$ far enough into the future that the normal congruence of geodesics has no conjugate points into the future. Let $t$  be proper-time on this congruence and let the space coordinates $x^i$ be comoving along it. Thinking of $e^{-Ht}$ as a defining function for $\scri$ one sees a resemblance between (\ref{s1},\ref{s2}) and the {\it{ambient metric construction}} of Riemannian geometers \cite{FG}. Using $\Theta=e^{-Ht}$ as conformal factor, we interpret $a_{ij}$ as the metric of $\scri$. It was shown in \cite{R1} that the metrics of Friedrich in \cite{F1,F2} can all be written in this form, but as far as I know this hasn't been proved for the metrics of \cite{LV, JTV,HS, S}, though it is a convenient way to present the results.

Clearly there is in the metric form (\ref{s2}) a freedom in the choice of initial surface, which effects the change
\[t\rightarrow\tt=t+f(x^i)+O(e^{-2Ht})\]
and requires a change of $x^i$:
\[x^i\rightarrow\tx^i=x^i-\frac{1}{2H}e^{-2Ht}a^{ij}f_j+O(e^{-4HT}).  \]
This in turn has the effect
\be\label{s3}(a_{ij},c_{ij})\rightarrow(\ta_{ij},\tc_{ij})=(\theta^2a_{ij},\theta^{-1}c_{ij})\ee
where $\theta=e^{-Hf}$, and this is the residual gauge freedom seen above and already noted in \cite{F1}.

 For all the examples we consider, the coefficient $b_{ij}$ has a universal expression
\be\label{s4}b_{ij}=H^{-2}(\rho_{ij}-\frac14\rho a_{ij})\ee
where $\rho_{ij},\rho$ are respectively the Ricci tensor and Ricci scalar of the metric $a_{ij}$.

Assuming that there is a Starobinsky expansion for the metrics of \cite{LV}, one can take from \cite{R1} that data on $\scri$ are $(a_{ij},c_{ij},M,U_i)$ with non-negative $M$, and expansions for the density and fluid velocity beginning
\be\label{s6}\mu=Me^{-4Ht}+O(e^{-5Ht}),\;\;u_i=U_ie^{Ht}+O(1),\;u^0=1+O(e^{-2Ht}),\ee
subject to constraints (\ref{s4}) and
\be\label{s5}a^{ij}c_{ij}=0,\;D_ic_j^{\;i}=-\frac{64\pi}{9H}MU_i.\ee
The rescaling freedom in this case is
\[(a_{ij},c_{ij},M,U_i)\rightarrow(\ta_{ij},\tc_{ij},\tilde{M},\tilde{U}_i)=(\theta^2a_{ij},\theta^{-1}c_{ij},\theta^{-4}M,\theta U_i),\]
as is clear from the expansions (\ref{s6}).

Again assuming there is a Starobinsky expansion for the Einstein-Vlasov metrics of \cite{JTV}, the data on $\scri$ become $(a_{ij},c_{ij},f(x^i,p_i))$ where $f$ is the non-negative distribution function (which doesn't change under rescaling). The constraints are that $c_{ij}$ is trace-free and satisfies the divergence condition
\[D_jc_i^{\;j}=-\frac{16\pi}{3H} \frac{1}{\sqrt{a}}\int fp_id^3p.\]

\section{Penrose's Weyl Curvature Hypothesis and data at the bang}

The Weyl Curvature Hypothesis is also due to Penrose, \cite{RP4}. He gives physical arguments for the initial singularity of the universe being very special as a singularity of a Lorentzian manifold, and conjectures that it is so special that while the space-time Ricci curvature is singular, the Weyl curvature is not: he conjectures that it must be finite or zero. It isn't immediately clear how to make a mathematical statement of this -- the metric and Ricci tensor are singular but the Weyl tensor is not -- but a simple strategy for doing so is to require that there is a conformal rescaling of the physical metric so that the initial singularity becomes a regular surface in an unphysical, extended manifold. An initial singularity for which this is possible has been variously defined as {\it{isotropic}} \cite{GW}, {\it{conformal}} \cite{N1,N2} or a {\it{conformal gauge singularity}} \cite{LT}. Given (\ref{i3}), if the unphysical Weyl tensor $C_{abc}^{\;\;\;\;\;d}$ is finite, as it would be in such a setting, then so too is the physical one $\tC_{abc}^{\;\;\;\;\;d}$ by (\ref{i3}), and we've imposed the Weyl Curvature Hypothesis..

It might be thought that the existence of such a conformal extension was a stronger assumption, even a {\emph{much}} stronger assumption, than finiteness of the Weyl tensor, but it is possible to show a local equivalence: given an incomplete conformal geodesic $\gamma$ ending at the singularity, and boundedness of the components of the Weyl tensor and its derivatives up to order $k+1$ in a Weyl-propagated frame along $\gamma$, there is a $C^k$ conformal extension of a terminal neighbourhood of $\gamma$ (see \cite{LT} for details).

It should be noticed that the conformal rescaling envisaged here has an important difference from that in (\ref{i1}): there the physical metric was becoming infinitely large towards the conformal boundary, so that $\Theta$ was going to zero at $\scri$; here, on the other hand, the physical metric is becoming very small (think what happens to the volume) so $\Theta$ is going to infinity. For that reason, it is easier to redefine the rescaling the other way round: suppose the physical metric $\tg_{ab}$ and unphysical metric $g_{ab}$ are related by
\be\label{4.1}\tg_{ab}=\Psi^2 g_{ab},\ee
with $\Psi=0$ at the conformal boundary, which we'll call $\Sigma$.

There is another importance difference which we may illustrate by considering FLRW metrics. Consider for simplicity the spatially flat FLRW metric
\be\label{4.2}ds^2=\tg_{ab}dx^adx^b=dt^2-(a(t))^2(dx^2+dy^2+dz^2),\ee
with a polytropic perfect fluid source, so that the energy-momentum tensor is
\[\tT_{ab}=(\rho+p)\tu_a\tu_b-p\tg_{ab}\mbox{   with  }p=(\gamma-1)\rho,\;\tu_adx^a=dt.\]
The usual range in $\gamma$ is from 1 (dust) to 2 (stiff matter) with 4/3 picked out as radiation and having a trace-free energy-momentum tensor. The conservation equation with the metric (\ref{4.2}) integrates to give
\[\rho=\rho_0a^{-3\gamma},\]
with $\rho_0$ a constant of integration, leaving the Friedmann equation as
\[\dot{a}^2=\frac{\kappa}{3}\rho a^2\]
with $\kappa=8\pi G$, to be solved. Thus
\be\label{4.3}a=a_0t^{2/3\gamma},\ee
choosing constants of integration so that $a=0$ at $t=0$, which is the location of the curvature singularity.
Conformal rescaling as in (\ref{4.1}) to add the boundary should clearly be done with $\Psi=a\sim t^{2/3\gamma}$ and this won't be differentiable at $\Sigma$ for any allowed 
$\gamma$. We can write it instead in terms of conformal time $\tau$, defined via $d\tau=dt/a$, obtaining
\[\Psi=a\sim \tau^{\frac{2}{3\gamma-2}}.\]
This is differentiable but with zero derivative at $\Sigma$ for $1\leq\gamma<4/3$, is smooth for $\gamma=4/3$ and not smooth for larger $\gamma$. This behaviour of the conformal factor is quite different from that in Section 3, where we chose it smooth or of some finite differentiability and with non-zero derivative at the boundary. 

The case of a radiation fluid, when $\Psi$ can be taken to be smooth, was considered first by Newman \cite{N1,N2}. He was able to reduce the Einstein equations for an irrotational radiation fluid with a conformal gauge singularity to a first-order system in a conformal time $\tau$ and comoving space coordinates, of the form
\be\label{4.4} A^0(u)\frac{\partial u}{\partial \tau}=A^i(u)\frac{\partial u}{\partial x^i}+B(u)u+\frac{1}{\tau}C(u)u\ee
with data $u(0)=u_0$. The coefficient matrices $A^0,A^i,B$ and $C$ are polynomials in the components of $u$, which include metric and connection components and gauge and constraint quantities, and with $B$ and $C$ zero, the system is symmetric hyperbolic, so $A^0,A^i$ are symmetric with $A^0$ positive definite. Such a system would more commonly be called {\emph{Fuchsian}} now. There is an obvious constraint on the data, that $C(u_0)u_0=0$ now commonly called the {\it{Fuchsian condition}}, but for his theorem, Newman requires an apparently stronger condition
\[C(u)u_0=0,\]
which in fact turns out to be no stronger. Newman also requires an eigenvalue condition, that the matrix $(A^0(u_0))^{-1}C(u_0)$ have no positive integer eigenvalues. From this Newman \cite{N2} proves well-posedness with data just the spatial 3-metric $h^0_{ij}$ of the data surface $\tau=0$. As a Corollary he notes that if the initial metric $h^0_{ij}$ is a metric of constant curvature then uniqueness of solutiion forces the solution to be FLRW: in this case the intial Weyl tensor is finite, but should it be zero initially then it will remain zero in the evolution.

Anguige and Tod \cite{AT1} were able to extend this result for $\gamma$ in the range $1<\gamma\leq 2$. In the extra cases, $\Psi$ is not smooth but one none-the-less arrives at a system like (\ref{4.4}) and concludes as before that the initial 3-metric $h^0_{ij}$ is all of the data -- the rescaled equations are well-behaved although the conformal factor is not. In particular one has the same Corollary: if the Weyl tensor is zero initially then it is always zero. This Corollary seemed unreasonably strong from a physical point of view, so consideration was given to massless Einstein-Vlasov solutions with a conformal gauge singularity, first with spatial homogeneity \cite{AT2} and later without \cite{A}. One still has well-posedness with data at the initial singularity but now the only datum is the initial distribution function $f^0(x^i,p_i)$ subject to a vanishing dipole condition. This in turn determines the initial 3-metric and fundamental form but in an indirect way. Now it is possible to have the Weyl tensor zero initially but becoming nonzero later.

Note we have zero $\Lambda$ in these results. One would expect the cosmological constant to have no effect near the initial singularity, just as one would expect replacing massless particles by massive ones in the Vlasov case to have negligible effect near the singularity, and this was verified at least for the spatially homogeneous case in \cite{T4}.

It was suggested \cite{AR} that one might find a bridge between the perfect fluid case and the Vlasov case, two cases for which the free data is notably different, by a consideration of the Einstein-Boltzmann equations, which might bridge the two extremes. With data at the initial singularity, this might as well be massless, and remarkably little seems to be known about the massless Boltzmann equation in curved space-time. However, in \cite{LNT} it was shown that for a certain set of reasonable scattering cross-sections, the isotropic and homogeneous case (i.e. FLRW) can be proved to be well-posed, and work-in-progress \cite{LNST} indicates that the same set of cross-sections give a well-posed system in Bianchi type I. The hard part is the Boltzmann equation!

\medskip

It should be noticed that the data required are very different for  the Cauchy problem with data given at the initial singularity as against data given at $\scri$: there is more freedom at $\scri$. This is essentially because, evolving back from $\scri$ one may have a solution become singular but it won't have the ordered nature necessary to be a conformal gauge singularity. Conversely, evolving forward from a conformal gauge singularity in the presence of a positive $\Lambda$ one is very likely to arrive at a decent $\scri$.

\section{Conformal Cyclic Cosmology or CCC}

We've seen that an expanding cosmological model with a positive $\Lambda$ is very likely to have a $\scri^+$, and then the conformal metric will extend through it and the Weyl tensor will vanish at it. We've also seen that a simple way to impose the WCH, of finite or zero Weyl curvature at the initial big bang, is to suppose that the conformal metric extends through the Bang surface. This will make the initial Weyl tensor finite rather than necessarily zero. It is hard, even impossible, to `cause' the WCH to hold \emph{from the future} but now there is a route to causing it \emph{from the past} and this is the essence of Penrose's outrageous suggestion: suppose that the conformal metric extends through both $\scri$ and the Bang surface, then $\scri$ can become the Bang surface of a subsequent {\emph{aeon}}, and at it the initial Weyl tensor is necessarily zero and the WCH holds. The physical metrics in the two aeons are different but are both conformally-related to a common, nonphysical `bridging' metric, and the Einstein equations are satisfied by both physical metrics. Each aeon is a complete universe evolving for an infinite proper time from an initial singularity, but lasting only for a finite conformal time before being followed by another aeon. There is no assumption of {\emph{periodicity}} but there is a conformal metric that extends through all aeons: the conformal metric can be said to be \emph{cyclic} whence Penrose's name for this model, Conformal Cyclic Cosmology or CCC.

\medskip 

The simplest model of this scenario, which we'll call \emph{the toy model}, is an FLRW metric with a radiation fluid as source and a positive cosmological constant. Take the metric to be
\[ds^2=dt^2-(a(t))^2d\sigma_k^2=a^2(d\tau^2-d\sigma_k^2),\]
where $a(t)$ is the scale factor, $d\sigma_k^2$ is one of the three standard constant-curvature Riemannian 3-metrics, $t$ is proper-time and $\tau$ is conformal time, so $dt=ad\tau$. The energy-momentum tensor is
\[T_{ab}=\frac43\rho(u_au_b-\frac14g_{ab}),\]
where $u_adx^a=dt$. From the conservation equation for this $T_{ab}$ we find that $\rho a^4$ is constant. Call this constant $\beta$ then the only remaining independent Einstein equation is the Freedman equation which in terms of conformal time is
\be\label{5.11}\left(\frac{da}{d\tau}\right)^2=\frac{\kappa\beta}{3}-ka^2+\frac{\Lambda}{3}a^4,\ee
where $\kappa=8\pi G$. We choose the solution with $a=0$ at $\tau=0$, so that there is an initial singularity, then $a(-\tau)=-a(\tau)$ and $a$ increases monotonically\footnote{We'll assume that $k\leq 0$ or $k=1$ and $4\kappa\beta\Lambda>9$, to avoid recollapse.} with $\tau$ and diverges to infinity at
\[\tau=\tau_F=\int_0^\infty\left(\frac{\kappa\beta}{3}-ka^2+\frac{\Lambda}{3}a^4\right)^{-1/2}da,\]
which is finite. Now it's straightforward to see from (\ref{5.11}) that
\[a(\tau_F-\tau)a(\tau)=\left(\frac{\kappa\beta}{\Lambda}\right)^{1/2},\]
so we can make a model for CCC by using $a(\tau)$ for $0<\tau<\tau_F$ and then replacing $a(\tau)$ by $a(\tau-\tau_F)=-\left(\frac{\kappa\beta}{\Lambda}\right)^{1/2}(a(\tau))^{-1}$ for $\tau_F<\tau<2\tau_F$, and so on: at each passage through $\scri$, $a$ is replaced by $c_1/a$ for suitable $c_1$. All aeons are diffeomorphic and satisfy the Einstein equations with the same $\rho$ and $\Lambda$.

It's a simple matter to add a dust source to this FLRW cosmology: add a term $\kappa\alpha a$ to the right hand side of (\ref{5.11}) where $\alpha$ is a constant of integration determining the dust contribution to the density as $\alpha a^{-3}$. One immediately loses the $a\rightarrow \mbox{constant}/a$  symmetry in the Freedman equation just noted but obtains a function $a$ more closely resembling that of the observed universe. Using values of $\alpha,\beta$ taken from observation, one can compute a more realistic $\tau(t)$ and as shown in \cite{T1} one can conclude that, if $\tau_0$ is the conformal time for `now' then the ratio $\tau_0/\tau_F$ is about 0.74: in other words, measured in conformal time the universe now is 74\% of the way to $\scri$. If we move on to the universe when it is ten times its present age (in proper time) then the fraction of conformal time remaining is just $10^{-4}$: the exponential expansion in proper time is really making itself felt, as the infinite proper time remaining is squeezed into a small amount of conformal time. For this reason, events occurring at times greater than $10t_0$ are essentially \emph{at} $\scri$.

\medskip

In a general model for CCC, we can suppose consecutive aeons have physical metrics $\hg_{ab}$ and $\cg_{ab}$ each conformally related to a single unphysical conformal metric $g_{ab}$ by conformal factors $\hOmega$ and $\cOmega$:
\be\label{5.12}\hg_{ab}=\hOmega^2g_{ab},\;\;\cg_{ab}=\cOmega^2g_{ab},\ee
with $\hOmega$ diverging at the common boundary $\scri$ and $\cOmega$ vanishing there. We assume that the product $\hOmega\cOmega$ is smooth and nonzero in a neighbourhood of $\scri$ and then we can change the choice of $g_{ab}$ so as to set\footnote{The minus sign is there because $\cOmega$ goes through zero at $\scri$; this equation holds where $\hOmega$ is finite, and as a limit at $\scri$.}
\[\hOmega\cOmega=-1.\]
This can't be done if some other process fixes the choice of $g_{ab}$ but otherwise it is just a gauge choice. It has the consequence that\footnote{Again this equation holds where $\hOmega$ is finite, and as a limit at $\scri$.}
\[\cg_{ab}=\cOmega^2g_{ab}=\cOmega^2\hOmega^{-2}\hg_{ab}=\hOmega^{-4}\hg_{ab}.\]
As a consequence the hatted and checked Ricci tensors have a relation like that in (\ref{i4}). Thus for example it won't generally be the case that both aeons can have simple perfect fluid sources -- at least one side has to have contributions like $\nabla_a\Upsilon_b$ to the Ricci tensor and therefore to the energy-momentum tensor: the toy model, where both sides are in fact perfect fluids, is seen to be exceptional. It's an outstanding question what the field equations should be in CCC and I'll make a suggestion for them below.

\medskip

CCC is intended to be a model of the actual universe so it must connect with observation. That isn't the principal interest of this meeting but it's worth touching on it briefly. First off, CCC automatically imposes the WCH. Next, an inevitable consequence of CCC is that the \emph{horizon problem} is dissolved: events widely separated on the sky may not have been causally connected in this aeon but they will have been if one takes account of earlier aeons. Thus we don't need inflation to solve the horizon problem and it is natural to try to do without it completely, at least near the Bang. If all that is needed physically for inflation is a period of exponential expansion, and since there will always be such a period late in any aeon, one may suppose that inflation happens \emph{before the Bang}, i.e. late in the previous aeon. In this way the well-known propensity of inflation to produce a suitable spectrum of density perturbations is retained.

Massless radiation, whether gravitational or electromagnetic, moves along null geodesics and so may be expected to pass through from one aeon to the following one. Late in the previous aeon the diversity of the universe is much reduced -- stars and galaxies will have gone and the universe will be populated by black holes which may be very large and will occasionally collide and merge, as well as evaporating by the Hawking process. These events will happen very close to $\scri$ in conformal time so the energy and radiation they generate will meet $\scri$ in a spherical annulus of limited radius or even a small ball. These regions of higher energy may be expected to have an effect on the mass-energy distribution on the last-scattering surface in the next aeon and therefore to give rise to characteristic imprints on the cosmic microwave background. There are in the literature claims to have detected these -- see references in \cite{ GP1,MN1} -- and other claims denying their significance -- see e.g. references in \cite{WE}. Roughly speaking, the arguments are over statistical significance. There is a need for a more detailed astrophysical account of the formation and energetics of the rings and the Hawking points, and indeed for an account of the transformation of physical quantities from one aeon to the next. It seems quite likely that there is a smoothing process between the data at the Bang from the previous aeon, which will mostly arise from a distribution of evaporating super-massive black-holes, and the last-scattering-surface in the next aeon, and repeated iteration of this smoothing may explain the homogeneity of the universe. It also seems quite likely that there is a place for inflation in the period of exponential expansion at the end of the previous aeon\footnote{This wsas suggested in \cite{RP3}.} and that this generates the density perturbations in the next aeon,  but these suggestions are speculative.

\medskip

I want to end with a suggestion of what the field equations of CCC might be, based on a paper of L\"ubbe, \cite{L}. Recall that a trace-free energy-momentum fits best with conformal methods, that one wants a component of the matter to have a radiation energy-momentum tensor, and that one needs some way to handle the $\nabla\Upsilon$ terms in (\ref{i4}), perhaps by having scalar fields in both aeons\footnote{Penrose \cite{RP3} has suggested that these terms be associated with a scalar field in the later aeon, proportional to a power of $\hOmega$, and that furthermore this field represents dark matter.}. Correcting slightly the account of \cite{L} that I gave in \cite {T1}, suppose we have a \emph{conformal scalar field}, that is a scalar field $\phi$ satisfying the \emph{conformal scalar field equation}
\be\label{5.2}Q(\phi, g_{ab},\alpha):=\Box\phi+\frac16R\phi-4\alpha\phi^3=0,\ee
where $\alpha$ is a constant to be chosen below. Such a scalar field has a kind of energy-momentum tensor given by
\be\label{5.1}D_{ab}(\phi,g_{cd},\alpha)=4\phi_a\phi_b-g_{ab}(g^{cd}\phi_c\phi_d)-2\phi\nabla_a\phi_b+2\phi^2L_{ab}+2\alpha\phi^4g_{ab}\ee
where
\[L_{ab}=-\frac12(R_{ab}-\frac16Rg_{ab}).\]
For the trace and divergence of $D_{ab}$ we calculate
\be\label{5.3}g^{ab}D_{ab}=-2\phi Q(\phi),\;\;\nabla^aD_{ab}=4Q\phi_b-2\phi Q_b,\ee
so that, given the field equation $Q=0$, $D_{ab}$ has the character of (plus or minus) a trace-free energy-momentum tensor for $\phi$, conserved by virtue of the field equation\footnote{Conformal scalars have been extensively used both in cosmology and in conformally-invariant extensions of the Standard Model of particle physics; see e.g. \cite{IS, IS2,MN,Sz}.}. Note $D_{ab}$ doesn't obviously satisfy any of the familiar energy conditions because of the term in $\nabla_a\phi_b$. Note also that, if $R=4\Lambda=24\alpha$ then $\phi=\pm 1$ are solutions (along with $\phi=0$ of course).

Under conformal rescaling
\[\tg_{ab}=\Omega^2g_{ab},\;\;\tphi=\Omega^{-1}\phi,\]
we have
\[D_{ab}(\tphi,\tg_{cd},\alpha)=\Omega^{-2}D_{ab}(\phi,g_{cd},\alpha),\;\;Q(\tphi,\tg_{cd},\alpha)=\Omega^{-3}Q(\phi,g_{ab},\alpha)\]
so the conformal scalar field equation is conformally invariant, and $D_{ab}$ transforms as an energy-momentum tensor should.

 If we regard $D_{ab}(\phi,g_{cd},\alpha)$ as the energy momentum tensor for $\phi$ subject to the field equation (\ref{5.2}) and include it in the Einstein equations with sources, then, choosing a sign, these can be taken to be
\[G_{ab}=D_{ab}(\phi)-\kappa T^{\mbox{ex}}_{ab}-\Lambda g_{ab},\]
where $T^{\mbox{ex}}_{ab}$ represents any other (`extra') matter fields present, assumed trace-free.

Now suppose $R=4\Lambda$ and $\alpha=\Lambda/6$ then $\phi=1$ is a solution of the field equation (\ref{5.2}) and we calculate from (\ref{5.1})
\[D_{ab}(1, g_{cd},\Lambda/6)=-G_{ab}-\Lambda g_{ab},\]
so the Einstein equations with these sources can be rewritten
\be\label{5.4}D_{ab}(1,g_{cd},\Lambda/6)+D_{ab}(\phi,g_{cd},\Lambda/6)-\kappa T^{\mbox{ex}}_{ab}=0.\ee
This is now conformally invariant: multiply it by $\Omega^{-2}$ and look at the three terms in turn:
\bean
\Omega^{-2}D_{ab}(1,g_{ab},\Lambda/6)&=&D_{ab}(\tphi:=\Omega^{-1},\tg_{cd},\Lambda/6),\\
\Omega^{-2}D_{ab}(\phi,g_{cd},\Lambda/6)&=&D_{ab}(\Omega^{-1}\phi,\tg_{cd},\Lambda/6),\\
\Omega^{-2} T^{\mbox{ex}}_{ab}&=&\tT^{\mbox{ex}}_{ab}.\eean
If we now choose $\phi=\Omega$ then the rescaling is
\[D_{ab}(\tphi,\tg_{cd},\Lambda/6)+D_{ab}(1,\tg_{cd},\Lambda/6)-\kappa \tT^{\mbox{ex}}_{ab}=0,\]
which is the same as (\ref{5.4}) but with tildes.

To make equations for CCC from this, suppose we have two conformal scalar fields, say $\phi$ and $\psi$, and the (bridging) metric $g_{ab}$ and consider the field equation
\be\label{5.5}D_{ab}(\phi,g_{cd},\Lambda/6)+D_{ab}(\psi,g_{cd},\Lambda/6)-\kappa T^{\mbox{ex}}_{ab}=0.\ee
This is conformally-invariant from what we have seen, so we can rescale to set $\phi=1$ (wherever $\phi$ is nonzero) giving $\hat{g}$  or to set $\psi=1$ (with the corresponding caveat) giving $\check{g}$. For the first, choose $\hat\Omega=\phi, $ so $\hg_{ab}=\hOmega^2g_{ab}$ and (\ref{5.5}) becomes
\[-\hG_{ab}-\Lambda\hg_{ab}+D_{ab}(\Phi,\hg_{cd},\Lambda/6)-\kappa\hT^{\mbox{ex}}_{ab}=0,\]
where $\Phi=\psi/\phi$, while for the second choose $\cOmega=\psi$ and $\cg_{ab}=\cOmega^2g_{ab}$ so that (\ref{5.5}) becomes
\[-\cG_{ab}-\Lambda\cg_{ab}+D_{ab}(\Psi,\cg_{cd},\Lambda/6)-\kappa\cT^{\mbox{ex}}_{ab}=0,\]
where $\Psi=\phi\psi^{-1}=\Phi^{-1}$. Thus the two aeons have the same field equations, just different fields. Following the suggestion of Penrose \cite{RP3} we may seek to interpret the conformal scalar fields $\Phi,\Psi$ as the dark matter in the respective aeons.

If this is going to be a model of CCC, we need $\phi$ to blow up at a  (space-like) surface $\scri$ and $\psi$ to be zero there, which can probably be arranged by choice of data (this is under investigation). We won't have $\hOmega\cOmega=-1$ unless $\phi\psi=-1$ at $\scri$ but that was a gauge choice we don't have to make.

\medskip

It remains to be seen whether CCC can be a viable theory of the universe we inhabit, but it fits very naturally into the story of conformal methods applied to cosmology.

\end{document}